# Realistic 3D printed imaging tumor phantoms for validation of image processing algorithms

Sepideh Hatamikia[1, 2, 3], Ingo Gulyas[4], Wolfgang Birkfellner[3], Gernot Kronreif[1], Alexander Unger[1], Gunpreet Oberoi[3], Andrea Lorenz[1], Ewald Unger[3], Joachim Kettenbach[5], Michael Figl[3], Janina Patsch[6], Andreas Strassl[6], Dietmar Georg[4], Andreas Renner[4]

[1]Austrian Center for Medical Innovation and Technology, Wiener Neustadt, Austria
[2]Research center for Medical Image Analysis and Artificial Intelligence (MIAAI), Department of Medicine, Faculty of Medicine and Dentistry, Danube Private University, Krems, Austria
[3]Center for Medical Physics and Biomedical Engineering, Medical University of Vienna, Vienna, Austria,
[4]Department of Radiation Oncology, Medical University of Vienna, Vienna, Austria
[5]Institute of Diagnostic, Interventional Radiology and Nuclear Medicine, Landesklinikum Wiener Neustadt, Wiener Neustadt, Austria

[6]Department of Radiology and Nuclear Medicine, Medical University Vienna

Corresponding author: Sepideh Hatamikia, sepideh.hatamikia@acmit.at, Viktor Kaplan-Straße 2/1, 2700 Wiener Neustadt

**Abstract**

Medical imaging phantoms are widely used for validation and verification of imaging systems and algorithms in surgical guidance and radiation oncology procedures. Especially, for the performance evaluation of new algorithms in the field of medical imaging, manufactured phantoms need to replicate specific properties of the human body, e.g., tissue morphology and radiological properties. Additive manufacturing (AM) technology provides an inexpensive opportunity for accurate anatomical replication with customization capabilities. In this study, we proposed a simple and cheap protocol to manufacture realistic tumor phantoms based on the filament 3D printing technology. Tumor phantoms with both homogenous and heterogenous radiodensity were fabricated. The radiodensity similarity between the printed tumor models and real tumor data from CT images of lung cancer patients was evaluated. Additionally, it was investigated whether a heterogeneity in the 3D printed tumor phantoms as observed in the tumor patient data had an influence on the validation of image registration algorithms.

A density range between -217 to 226 HUs was achieved for 3D printed phantoms; this range of radiation attenuation is also observed in the human lung tumor tissue. The resulted HU range could serve as a lookup-table for researchers and phantom manufactures to create realistic CT tumor phantoms with the desired range of radiodensities. The 3D printed tumor phantoms also precisely replicated real lung tumor patient data regarding morphology and could also include life-like heterogeneity of the radiodensity inside the tumor models. An influence of the heterogeneity on accuracy and robustness of the image registration algorithms was not found.

## 1) Introduction

Medical imaging offers a wide variety of imaging modalities to gain information required for diagnosis or therapy control. New imaging algorithms need extensive testing and validation. However, the ground truth for validation of different imaging algorithm is hard to obtain. Thus, sophisticated phantom studies are required in order to precisely evaluate the performance of newly developed algorithms. For CT imaging besides realistic morphology, the radiological properties also need to be simulated accurately. The availability of additive manufacturing (AM) technology, colloquially called 3-dimensional (3D) printing, has paved the way towards creation of realistic models and imaging phantoms for specific research purposes in many fields as well as in radiation therapy (RT) [1-5]. CT-derived 3D printed anatomical models allow for creating patient-equivalent structures with lifelike shape and internal structure [1, 5]. So far, several phantoms have been developed to mimic bone and soft tissue related to different site in the body [6-9]. However, the possibility of developing sophisticated phantoms that mimic both, the X-ray attenuation properties and the morphology of the patient's tumor tissue, has not yet been extensively investigated. Realistic tumor phantoms are required for validation of different image processing algorithms such as registration algorithms (e.g., tumor tracking for tumor motion monitoring in RT) and image segmentation (e.g., validation of advanced automatic segmentation methods). Based on our knowledge, there is no study describing a methodology suitable to manufacture realistic tumors with both human-like morphology and radiodensity suitable for CT imaging. In addition, the influence of tumor-like heterogeneous radiodensity distributions in imaging phantoms was not investigated in the previous literature. Thus, the question whether a simplification introduced by the imaging phantom (e.g homogenous radiodensity other than heterogenous) in a validation study affects the accuracy of the developed imaging algorithm could not be answered.

Some studies have tried to replicate realistic radiation attenuation properties using different 3D printer-based technologies. In a study [10], a life-size thorax phantom including different tumors was developed using 3D printing technology based on a clinical CT scan from a lung cancer patient in order to closely resemble patient structure and tissue density. Three lung tumors were printed using nylon material based on Selective Laser Sintering (SLS) technique. Although their 3D printed results showed a comparable result with patient CT regarding the shape, size and structure, the Hounsfield units (HU) were not similar to those of the patient CT data. In another study [11], authors used Acrylonitrile Butadiene Styrene (ABS)-M30 thermoplastic material to print a tumor replica based on a non-small cell lung cancer (NSCLC) patient CT data. Their 3D printed tumor was used to investigate dosimetric experiments and imaging purposes using an advanced breathing phantom (ARDOS). Although authors claimed that their tumor phantom provided a realistic shape and HU parameters, they did not report the radiodensity value of the manufacture's tumor phantom and its correspondence to the real patient data. In another study [12], agar-gelatin-based tumor phantoms were developed which could both visually and radiographically mimic typical head and neck squamous cell carcinoma (HNSCC). Tumor phantoms were used in a cadaver model and a radiodensity between 59 and 127 HU (mean 93.7) was achieved which was in a good agreement with the patient tumor radiodensity. However, the inserted tumor could not mimic a realistic tumor shape. In addition, their proposed tumor phantom does not mimic the density heterogeneity that is seen with most HNSCC [12].

In this study, we propose a simple protocol for manufacturing human-like imaging tumor phantoms regarding morphology and radiodensity using filament printing technology. Different filament materials were investigated in order to find their correspondence with realistic radiodensity range in patient tumor data and a radiodensity spectrum with a good match to the lung patient tumor radiodensity range was explored. In addition, a look-up table with a HU range between -217 and 226 was introduced to serve as a recipe for researchers to create their individual customized CT tumor phantoms (for different types of tumors) with the desired range of radiodensities. As an additional analysis, 3D-printed realistic tumor phantoms capable of simulating the heterogeneous structure and radiological characteristics of tumors found in CT of lung cancer patients were studied and the effects of using heterogeneous tumor phantom compared to a homogeneous tumor phantom on the performance of the image registration algorithms were evaluated.

### 2) Materials and Methods

**2.1) Patient data**

In this study, we investigated the CT data from seventeen lung cancer patients. The data used in this study was approved by the Ethics Committee of the Medical University of Vienna (EK1253/2012). Anonymized patient CT data (SOMATOM Definition AS, Siemens Healthineers, Germany, tube voltage 120 kVp, tube current time product 315 mAs) including Digital Imaging and Communication in Medicine (DICOM) data files was used.

**2.1.1) Determination of realistic HU range in patient tumor CT data**

The CT from different patient tumor data showed large differences in the shape and size. Five examples of the patient tumor data with different sizes are shown in Fig. 1a. We evaluated the radiation attenuation property of the tumors by determining HU values corresponding to the segmented CT data. A HU spectrum analysis was performed in order to quantify the HU distribution of the tumor voxels. Furthermore, in order to investigate HU inhomogeneity (or distribution) within the tumor, the Segmentations module from 3D Slicer software 4.10.2 (Salt Lake City, Utah, USA) was used to segment the tumor according to different HU ranges for identifying regions with specific HU ranges. The HU analysis (Section 2.7) using all patient tumors demonstrated a HU range between -100 to 100 for most of patient tumors. In addition, a higher radiodensity (an average range between 0 to100 HU) in the centroid of the tumor while lower radiodensity (an average range between -100 to 0 HU) in the outer areas of the tumor was observed which showed a HU heterogeneity within different areas inside the tumor. Visualization of a HU spectrum and the corresponding tumor segmented regions for one example of patient data is illustrated in Fig.1 b-f.

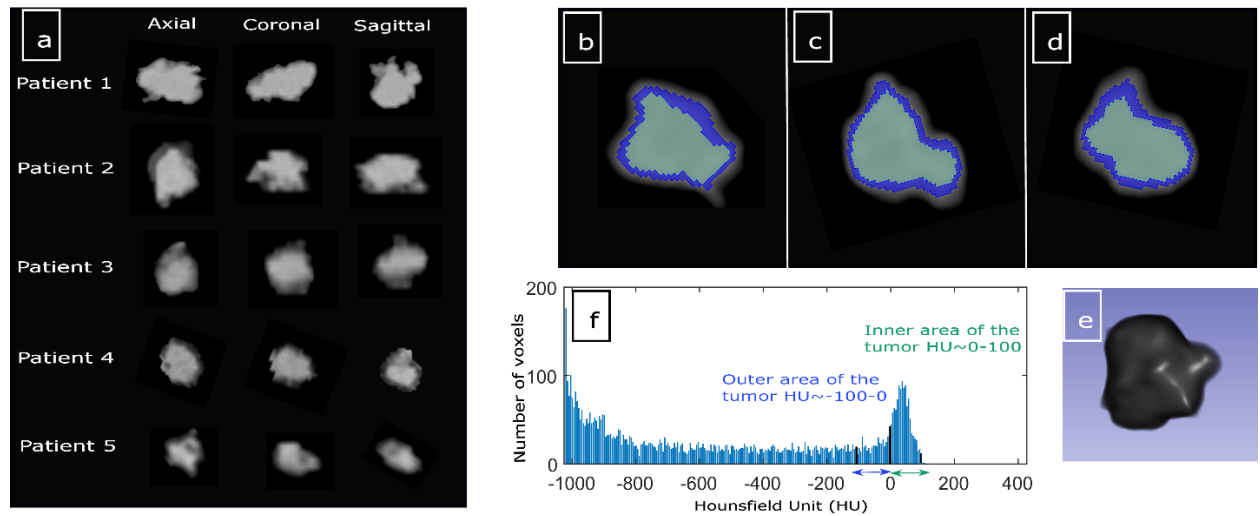

**Figure.1.** (a) Five examples of the patient tumor CT data with different sizes, CT segmentation of one sample tumor (patient 4) in three different views (b-d) as well as 3D representation in 3D Slicer (e) and the corresponding HU

spectrum (f) in order to visualize the HU inhomogeneity within the tumor. The spectrum in the range of -1000 to -200 HU is related to surrounding lung tissue.

## 2.2) 3D Phantom Design

SlicerRT module in 3D Slicer was used for designing the tumor phantoms. The segmentation of the clinical target volume (CTV) was used to extract the tumor volume from the CT patient data in order to create the tumor models. The tumor segmentation was exported as STL (Standard Tessellation Language) file to extract the smoothed tumor surface used as input data for the PrusaSlicer Software 2.3.0 (Prusa Research a.s., Praha, Czech Republic) for the 3D printing procedure. For the heterogenous tumor models, two separate models were created from the segmented inner and outer areas of the tumors (e.g. segmented areas as shown in Fig.1 b-d). In this study, two different tumor models with different tumor volume were used: tumor model 1 originating from the patient 1 CT data and tumor model 2 originating from the patient 4 CT data (Fig. 1a). A 3D visualization of designed tumor models after rendering in Autodesk Meshmixer software 3.5.474 (Portland, Oregon, USA) software is shown in Fig. 2 a, b.

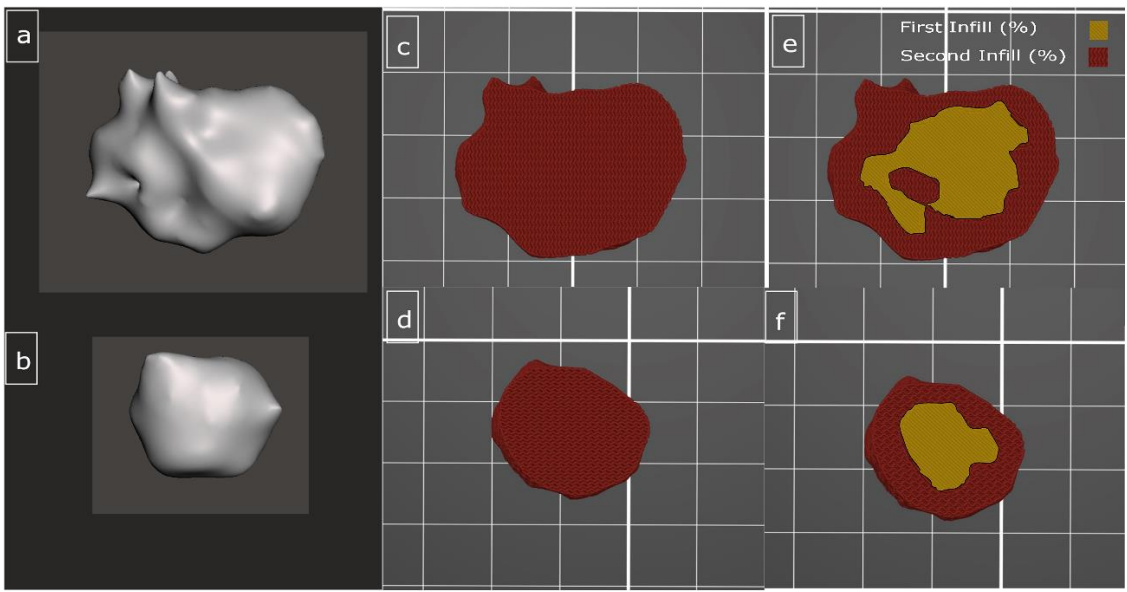

**Figure 2.** (a, b) 3D visualization of the tumor models in Autodesk Meshmixer software. Illustration of the design of the homogenous (c, d) and heterogenous (e, f) for both tumor models in the PrusaSlicer software. Difference between infill densities used for the inner area and outer areas of hetrogenous tumor is visible (e, f).

## 2.3) Additive manufacturing using filament printing technology

Using Fused Deposition Modeling (FDM) technology, a 3D object is printed layer by layer by extruding a filament through a nozzle. The filament can be printed in a wide variety of grid patterns, which is called infill, and the infill density (material:air ratio) can be set in %. Using this technique, prints even with the highest infill density might include air gaps in the sub-millimeter scale. This feature makes this technique suitable to manufacture realistic imaging phantoms, which are able to simulate a grey value spectrum equivalent to the radiation attenuation of human tissues by means of varying the infill density parameter.

In a study [13], the radiological properties of a variety of 3D printing materials were evaluated for different ranges of photon energies. Using FDM printer, the ABS and Polylactic Acid (PLA) filaments were printed with variety of infill densities to achieve different CT values approximating the CT values of low-density lung, high-density lung and soft tissues. In another study [14], authors also investigated a large set of commercially available 3D-printed samples from different plastic materials to evaluate their radiological properties. The main focus was to find suitable tissue-equivalent 3D-printed plastics materials for medical phantoms which are used in radiology. In this study, we investigated different FDM-based filament materials and finally introduced the most promising materials with suitability for mimicking lung tumor radiation attenuation properties as observed in lung cancer patient datasets.

**2.4) Creation of cylindric samples with specific radiation attenuation properties**

In order to determine the infill densities that reproduce different radiodensities comparable with the patient tumor CT we evaluated a large set of different filament materials and varied the infill density parameter of the print. A cylinder model with the diameter of 2 cm and length of 7 cm was designed for the test printing. For each filament material, a set of 7 infill densities including 100%, 97%, 94%, 91%, 88%, 85%, 82% (called as Infill 1, Infill 2, ... Infill 7) was used to reproduce a gradient in HU range within a cylinder. Four different filament materials including Polylactic acid (PLA), Polayamid 12 (Nylon12 or PA12), Acrylonitrile styrene acrylate (ASA Pro), Polyethylenterephthalat (PETG) were selected and reported in this study due to their similarity to the patient tumor radiodensity range. In addition, a set of 22 infill densities including 100% to 99% with infill step size of 1% was used for two additional cylinders (PLA and PET) with similar cylinder size as before. This experiment was done to investigate whether a continuous change of the HU can be achieved within the printed model.

**2.5) Density values selected to be replicated in the tumor phantom**

For each tumor model (Section 2.2), three heterogenous and three homogenous samples with different infill densities were designed. The selection of infill densities for the samples was done after matching the resulted HU values achieved from the cylinders CT with the corresponding CT values in the patient tumor scans. Different infill densities were used for different tumor samples in order to simulate different HU range as observed in the patient dataset (Table 1).

**Table1.** Infill densities and materials used for all samples for tumor models 1 and 2

|  | Sample 1 | Sample 2 | Sample 3 |
| --- | --- | --- | --- |
| Homogenous tumor model 1 | Material: PLA<br>Total area: Infill 7 | Material: PLA<br>Total area: Infill 3 | Material: PLA<br>Total area: Infill 4 |
| Homogenous tumor model 2 | Material: Nylon<br>Total area: Infill 3 | Material: Nylon<br>Total area: Infill 1 | Material: Nylon<br>Total area: Infill 2 |
| Heterogenous tumor model 1 | Material: PLA<br>Inner area: Infill 3<br>Outer area: Infill 7 | Material: PETG<br>Inner area: Infill 1<br>Outer area: Infill 5 | Material: ASA<br>Inner area: Infill 1<br>Outer area: Infill 2 |
| Heterogenous tumor model 2 | Material: PLA<br>Inner area: Infill 2<br>Outer area: Infill 7 | Material: PETG<br>Inner area: Infill 1<br>Outer area: Infill 6 | Material: ASA<br>Inner area: Infill 1<br>Outer area: Infill 2 |

**2.6) Additive manufacturing of the tumor phantoms**

The 3D printing was done using a custom filament 3D printer (Original Prusa i3 MK3S, Prusa Research a.s., Praha, Czech Republic). The G-Code for printing was prepared with the PrusaSlicer Software. A Gyroid pattern was used for the infill. Infill density was adapted to match the respective HU values. The layer resolution was set to 0.2 mm for all samples. Printing temperature and speeds were selected based on the filament manufacturer and 3D printer specifications. In order to define different infill densities for inner and outer areas for heterogenous tumors in PrusaSlicer (Fig. 2 e, f), the whole STL model was imported as one part and different sub-parts were added as modifiers.

**2.7) HU analysis of the patient CT, tumor phantom and test samples**

The test printed cylinders as well as tumor phantoms were scanned in a CT (Section 2.1) with the standard clinical CT protocol (tube current time product 315 mAs, tube voltage 120 kVp, slice thickness 2 mm, pixel spacing 0.68 mm) in order to evaluate their density properties. We used the Analyze 12.0 toolkit (AnalyzeDirect, Overland Park, United States) in order to measure the HU values from the CT scan related to patients, printed cylinders and printed tumor phantom. The HU analysis was done by selecting different line profiles inside the corresponding region and measuring the HU by calculating the average and the standard deviation over all points for the selections related to those line profiles.

**2.8) Physical dimensional comparison between the phantom and patient STL**

The FDM printed tumor samples were individually scanned using a high-resolution CT scan (Syngo CT VA40A, Siemens Healthineers, Germany, tube current time product 200 mAs, tube voltage 120 kVp, slice thickness 0.2 mm, pixel spacing 0.09 mm) in order to evaluate their geometry accuracy as compared to the original tumor model. The resulted DICOM files were exported to the segmentation and modelling software Materialize Mimics Research 23.0 (Materialize, Leuven, Belgium) for a semi-automatic thresholding of the tumor samples. All samples were segmented and exported to the designing software Materialize 3-Matic 15.0 for dimensional comparison (Fig. 3). The software gave volume and surface area information of the samples and patient tumor models (Table 2). Both manual alignments followed by automatic N-point registration of the printed tumor models over the patient tumor models was performed by a single operator, and overlapping volume was calculated manually and as well as using Collision detection analysis, for each sample. The latter detects intersections that occur between multiple parts and calculates the colliding volume.

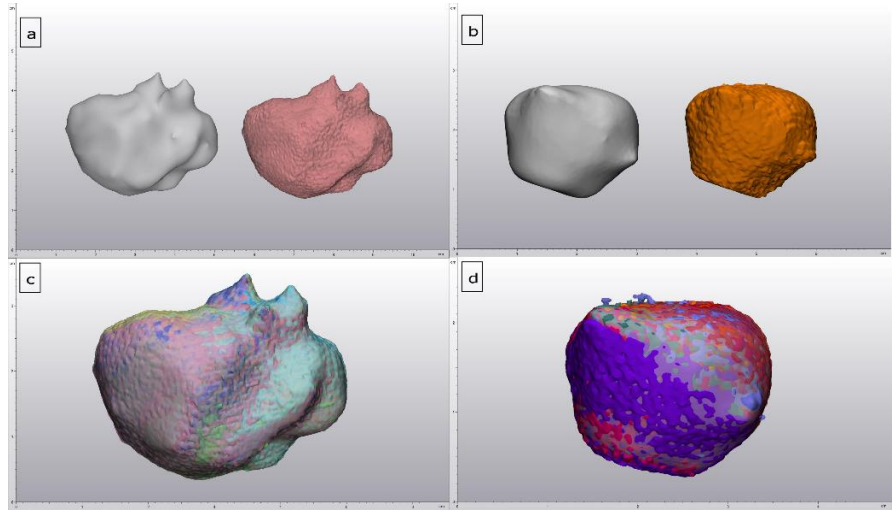

**Figure 3.** Dimensional analysis workflow in 3-Matic 15.0 software. a) Digital models of patient (grey) and 3D printed tumor (pink) for tumor model 1, b) Digital models of patient (grey) and 3D printed tumor (orange) for tumor model 2, c) and d) Manual followed by automatic registration of six 3D printed tumor samples on the patient tumor models for tumor model 1 (c), tumor model 2 (d). The colors represent the different tumor sample models.

**Table 2**. Table showing the volume and surface area of the 3D printed tumor samples (samples 1-6 are related to both heterogenous and homogenous models), generated automatically in 3-Matic 15.0 software. This is directly based on the segmentation of the tumor CT images in Mimics Research 23.0.

|  | Volume of tumor model 2 (mm3) | Surface area of tumor model 2 (mm2) |
|---|---|---|
| **Patient** | **5.344,81** | **1.882,12** |
| **Sample 1** | 4.369,25 | 1.621,34 |
| **Sample 2** | 4.052,86 | 1.382,68 |
| **Sample 3** | 4.039,88 | 1.384,37 |
| **Sample 4** | 4.211,99 | 1.395,93 |
| **Sample 5** | 4.363,74 | 1.436,14 |
| **Sample 6** | 4.389,59 | 1.508,97 |
|  | **Volume of tumor model 1 (mm3)** | **Surface area of tumor model 1 (mm2)** |
| **Patient** | **11.611,15** | **2.833,45** |
| **Sample 1** | 11.480,50 | 2.997,60 |
| **Sample 2** | 11.477,12 | 2.978,05 |
| **Sample 3** | 11.551,95 | 2.954,48 |
| **Sample 4** | 11.633,25 | 3.328,17 |
| **Sample 5** | 11.402,31 | 3.154,86 |
| **Sample 6** | 11.666,98 | 3.056,08 |

## 2.9) Breathing phantom for image registration studies

To obtain a ground-truth tumor motion we used ARDOS phantom [11]. The phantom has a removable rib cage and two lungs made from balsawood. Balsawood has the advantage of having low HUs as well as being a heterogeneous material with a HU distribution similar to lung tissue. Both the rib cage and the whole body including the lungs can be moved in axial direction to mimic breathing motion. Additionally, each lung has a borehole that enables to place inserts, which can be moved independently from the rest of the phantom. The inserts (Figure 4b) are also made from balsawood and can hold tumors of different size. A holder cast from silicone (Figure 4c) was constructed for reproducible positioning of the tumor phantoms inside the balsawood inserts (Figure 4d, e).

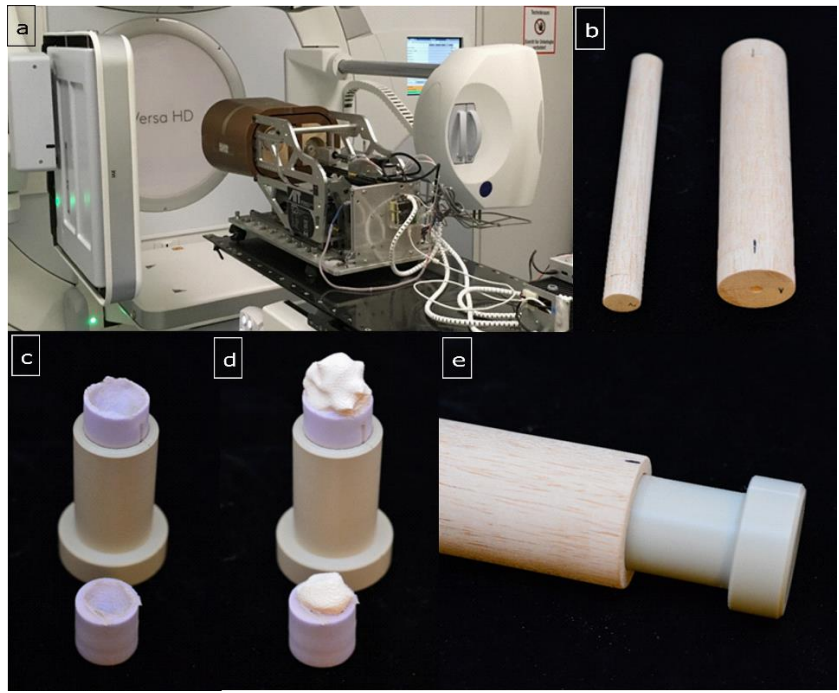

**Figure 4.** The ARDOS breathing phantom (a) and the inserts which can hold printed tumors (b). A holder (c) was build for each of the tumor models with the exact imprint of the tumor to hold it in position (d) for reproduceible placement inside the insert (e).

## 2.10) Measurements

3D CTs from the phantom were acquired at three positions of the tumor inserts simulating a breathing motion amplitude of 24 mm:
- position 1) the initial position simulating the maximum inspiration position,
- position 2) the tumor was moved forward 12 mm in longitudinal axis simulating a respiratory center position,
- position 3) the tumor moved another 12 mm forward in longitudinal axis simulating maximum expiration.

All CTs were acquired both with and without the rib cage placed inside ARDOS for every 3D printed tumor sample. Additionally, 2D projective x-ray images were acquired using the Elekta XVI imaging system in posterior-anterior (PA) and lateral (LAT) direction and again for every 3D printed tumor sample both with and without the rib cage placed inside ARDOS. The simulated breathing motion amplitude was again 24 mm but this time with steps of 2 mm leading to 13 imaging positions.

### 2.11) 3D/3D image registration

In order to perform the 3D/3D registration, we used the 3D Slicer toolbox using the General Registration (Elastix) module. In this study, for each tumor, the 3D registration was performed between CT images after two movements: 1) between images acquired at position 1 and position 2 (first motion), 2) between images acquired at position 2 and position 3 (second motion) (Section 2.10). The 3D/3D registration error related to CC, LAT and PA directions as well as a 3D error by calculating the Root Mean Square Error (RMSE) over the errors from these three directions were calculated.

### 2.12) 2D/3D image registration

For 2D/3D registration the fast intensity-based image registration (FIRE) was used. FIRE is an in-house developed software for real time 2D/3D registration [15, 16]. It uses digitally reconstructed radiographs (DRRs) generated from CT data which are registered to x-ray images. For the present study tumor registration was performed with an optimizer setting of 2 degrees of freedom (DoF). All 2D images (13 images with a motion step of 2 mm per measurement set) were registered to the CT in position 1) (Section 2.10).

## 3. Results

### 3.1) Resulting HU values for the printed cylinders

The four test cylinders with different infill density (Section 2.4) were printed successfully (Fig. 5 a). The CT scan from the four cylinders is presented in Fig. 5 b, c. A range between -217 to 226 HUs was achieved using four different materials (Table 3).

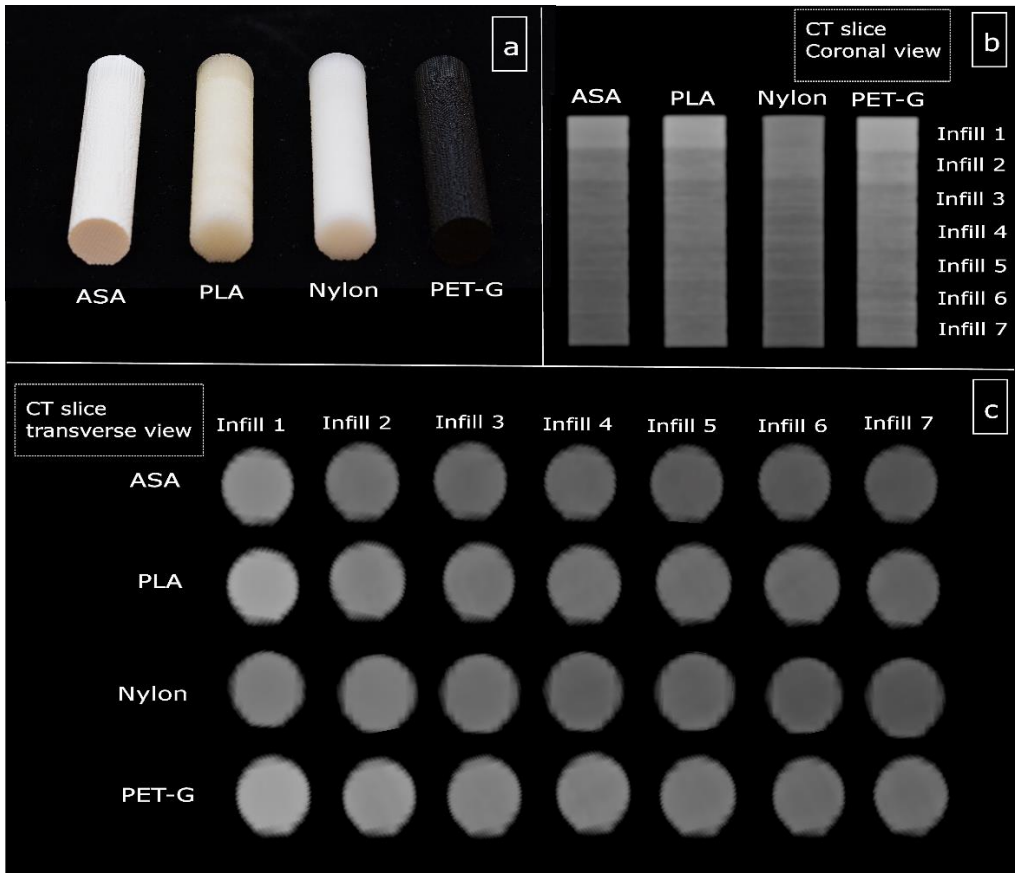

**Figure 5.** (a) The printed cylinders with four different materials including ASA, PLA, Nylon and PETG. Coronal and transverse views of the CT image from four printed cylinders at different infill densities.

**Table 3.** HU values related to all cylinders from four different materials including ASA, PLA, Nylon and PETG at different infill densities. Std: standard deviation.

|  | Infill % | ASA (HU±Std) | PLA (HU±Std) | Nylon (HU±Std) | PETG (HU±Std) |
|---|---|---|---|---|---|
| **Infill 1** | 100 | 155.30±8.66 | 226.78±14.21 | 50.14±8.34 | 245±8.41 |
| **Infill 2** | 97 | 19.13±8.71 | 98.40±10.76 | 18.43±5.80 | 180.52±14.90 |
| **Infill 3** | 94 | -38.10±8.10 | 65.72±14.68 | -56.12±13.10 | 144.38±16.36 |
| **Infill 4** | 91 | -63.72±17.28 | 7.23±6.90 | -93.36±10.12 | 117.68±14.54 |
| **Infill 5** | 88 | -105.85±13.04 | -24.57±8.16 | -121.20±13.41 | 85.39±18.17 |
| **Infill 6** | 85 | -130.79±16.74 | -47.85±20.56 | -185.34±15.24 | 39.05±13.61 |
| **Infill 7** | 82 | -150.21±14.86 | -85.29±11.60 | -217.32±14.10 | -10.44±7.63 |

### 3.2) Additively manufactured tumor phantoms

All tumor phantoms from the two tumor models (Fig. 2 a, b) were successfully printed using the FDM printer. (Fig. 6).

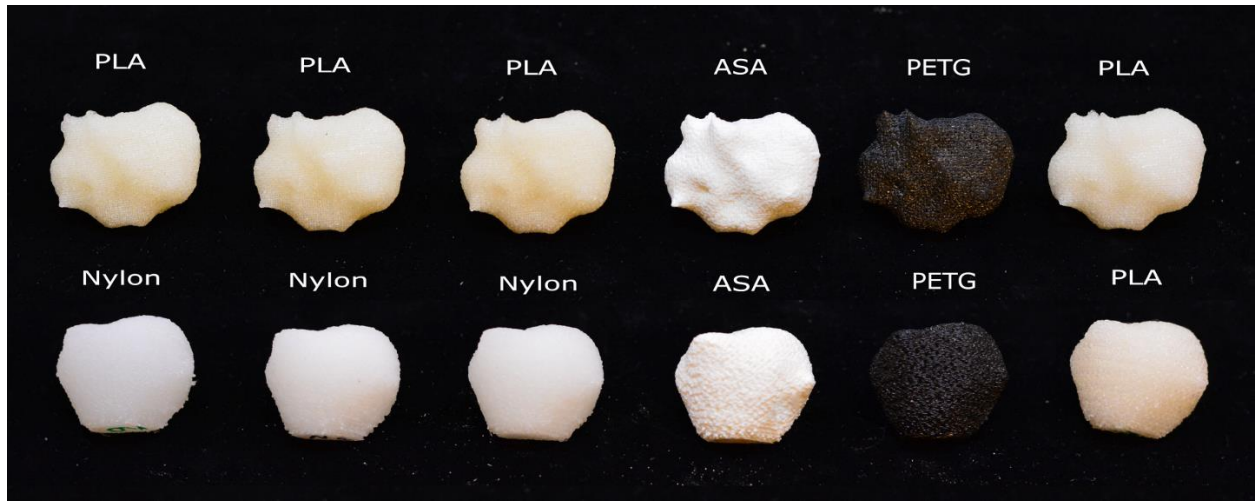

**Figure 6.** The 3D printed tumor phantoms using four different materials including PLA, ASA, PETG and Nylon.

### 3.3) Resulting density values for the tumor phantom

An axial slice from the patient and phantom CT are shown in Fig. 7. The printed heterogenous tumor sample 1 for model 1 showed an average density of 62 and -64 HU for inner and outer areas of the tumor (Table 4). In addition, the printed heterogenous tumor sample 1 for model 2 showed an average HU of 64 and -53 for inner and outer areas of the tumor. These results showed a good agreement between different density values of the proposed tumor phantom and the initial patient CT scan (average HUs of 0 to100 (=50) HU for inner area and average HU of -100 to 0 (=-50) for outer area) for the corresponding areas. The printed homogenous tumor sample 3 for model 1 and model 2 showed an average HU of 9 and -3 respectively. Additionally, these results showed a good match between the standard tumor phantom HU and the average patient tumor of -100 to 100 (=0) HU. The other tumor samples were designed to represent also other HU distribution ranges as observed in other patient tumor data (patient data which had some deviation from the average HU (-100 to 100)).

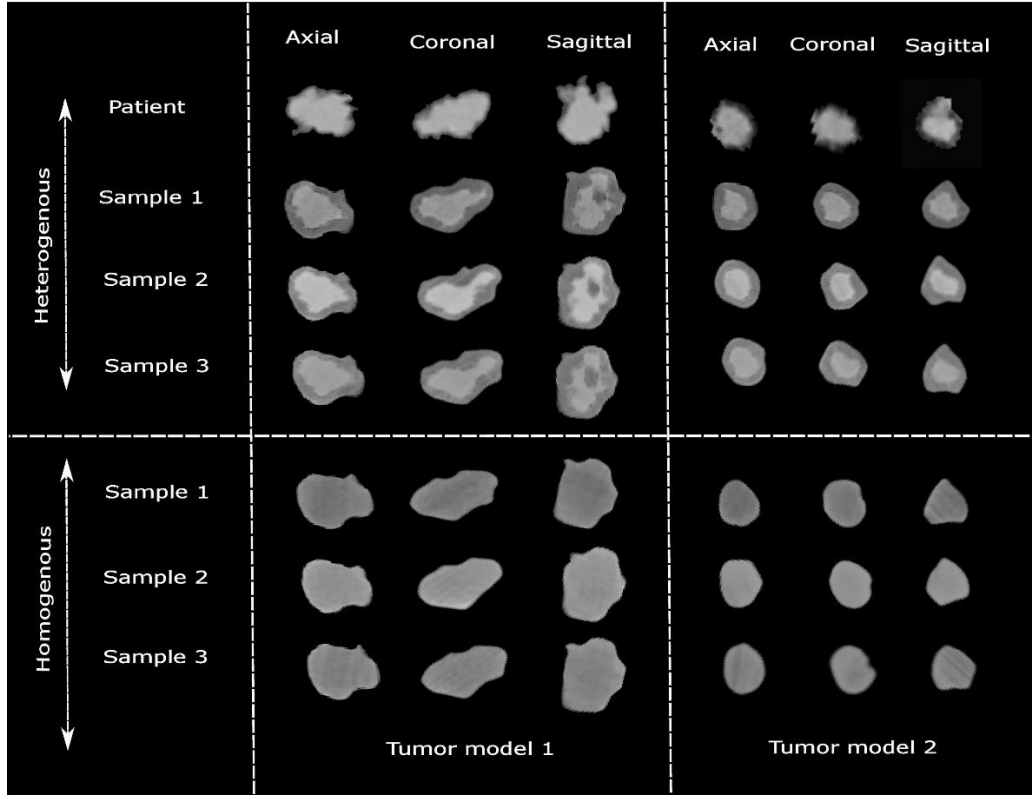

**Figure 7.** An axial slice from the patient and tumor phantom CT for different tumor models and samples.

**Table 4.** HU values related to all printed tumor phantoms. Std: standard deviation.

|  | Sample 1 (HU±Std) | | Sample 2 (HU±Std) | | Sample 3 (HU±Std) | |
|---|---|---|---|---|---|---|
| **Heterogenous tumor model 1** | **Inner area** 62.43±14.12 | **Outer area** -64.17±9.32 | **Inner area** 233.56±7.51 | **Outer area** 78.86±24.78 | **Inner area** 138.12±6.79 | **Outer area** 28.80±9.87 |
| **Heterogenous tumor model 2** | **Inner area** 64.64±9.83 | **Outer area** -53.40±18.12 | **Inner area** 227.31±10.28 | **Outer area** 46.83±16.17 | **Inner area** 152.18±4.88 | **Outer area** 29.45±6.60 |
| **Homogenous tumor model 1** | **Total area** -74.80±10.32 | | **Total area** 85.72±12.30 | | **Total area** 9.53±13.58 | |
| **Homogenous tumor model 2** | **Total area** -63.33±7.46 | | **Total area** 53.94±6.31 | | **Total area** -3.64±25.35 | |

### 3.4) 3D/3D registration results

According to the results, the average 3D RMSE over the three samples and different motions for the heterogenous tumor model 1 and tumor model 2 were 0.119 and 0.126, respectively. In addition, the average 3D RMSE over the three samples and the two motions for the homogenous tumor models1 and tumor model 2 were 0.102 and 0.108, respectively. These results show that for tumor model 1, the difference between 3D RMSE of heterogenous and homogenous design was 0.017. Furthermore, for tumor model 2, the difference between 3D RMSE of heterogenous and homogenous design was 0.018.

**3.5) 2D/3D registration results**

For PA imaging the 2D/3D registration was performed in CC/LAT directions. In addition, for LAT imaging the registration was done in CC/PA direction for both tumor models. The registration was successful on all x-ray images and total 2D errors of 0.043 mm and 0.035 mm was found for experiments without ribs for homogenous and heterogenous tumor samples, respectively (difference of 0.008 mm). In addition, total errors of 0.594 mm and 0.539 mm were found for the analysis including ribs for homogenous and heterogenous tumor samples, respectively (difference of 0.055 mm).

**3.6) The results from the physical dimensional comparison between the phantom and patient STL**

The results showed that the printed volume of the tumor, was lower respectively than that of the patient by 0.65 % for the tumor model 1 and 20.71% for the tumor model 2 and the surface was rough with overall indentations; the surface area automatically generated by the 3-Matic 15.0 software was 8.64% higher and 22.7% lower than the patient model. During Collision Detection, the resulting volume overlap was 96.72 % for the tumor model 1 and 78.15% for the tumor model 2 (Table 5).

**Table 5.** Table showing the colliding volume between the 3D printed models (samples 1-6 are related to both heterogenous and homogenous models) and patient tumor model per sample, generated by Collision Detection function in 3-Matic 15.0 software.

| Model | Colliding volume of tumor model 1 (mm3) | Colliding volume of tumor model 2 (mm3) |
|---|---|---|
| **Sample 1** | 11.291,33 | 4.243,47 |
| **Sample 2** | 11.076,82 | 4.048,21 |
| **Sample 3** | 11.245,25 | 4.034,05 |
| **Sample 4** | 11.299,77 | 4.179,00 |
| **Sample 5** | 11.115,99 | 4.305,23 |
| **Sample 6** | 11.355,22 | 4.251,84 |
| **Mean colliding volume** | **11.230,73** | **4.176,97** |
| **STDEV** | 110,45 | 112,68 |
| **% overlap** | 96,72 | 78,15 |
| **% failure** | -3,28 | -21,85 |

**3.7)** Resulting HU values for the printed cylinders with continuous change of the HU

The CT scan from the two cylinders including 22 infills with step size 1% is presented in Fig. 12. Ranges between -132 to 230 HUs and -65 to 256 HUs were achieved for PLA and PETG materials, respectively.

The CT result (Fig. 8) shows that a continuous change of the HU can be achieved when using step size of 1%.

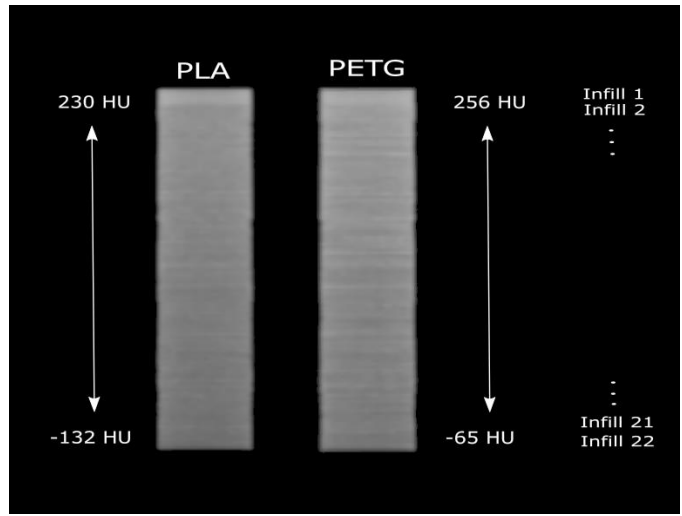

**Figure 8.** (a) The printed cylinders with four different materials including ASA, PLA, Nylon and PETG. Coronal views of the CT image from two printed cylinders (PLA and PETG) at different infill densities with step size of 1%.

### 4) Discussion

In this study, a methodology for fabrication of CT-derived 3D printed tumor phantoms with realistic radiodensity and morphology were described. Four top materials qualified for mimicking the clinically relevant HU range for tumor tissues were introduced. Additionally, the infill densities of the materials were modulated in small step sizes to achieve a gray value spectrum equivalent to human lung tumor tissue (Table 2). The resulting gray value spectrum with the corresponding infill density percentages reported in this study represents a lookup-table for researchers and phantom manufactures to create realistic CT tumor phantoms with the desired range of radiodensities. To our knowledge this is the first study which proposed direct fabrication of realistic tumor phantoms. The only two previous studies focusing on the development of imaging tumor phantoms, reported limitations regarding replicating either the density or the structure compared to real patient tumor data [10, 12]. In [10] CT-derived tumor phantoms using SLS technique were 3D printed and found to be geometrically highly accurate compared to real patient lung tumor data. However, their tumor phantoms could not replicate the realistic radiodensity of tumors. In another study [12], although the proposed tumor phantoms density was in a good agreement with the patient tumor radiodensity, a realistic tumor shape and structure could not be reproduced using their approach based on injection of agar-gelatin. Our proposed design enhances previous studies [10, 11, 12] by direct manufacturing of solid imaging tumor phantoms which can closely resemble patient tumor data regarding both radiodensity and the geometry. Furthermore, this is the first study which proposes a phantom design that facilitates simulating complex density heterogeneity inside the tumor phantoms. In heterogenous tumor models, we selected only two infill densities, one for inner and one for outer volume which resulted to distinct HU difference. However, our experiments using some additional cylinder models also showed that a continuous change in HU can be achieved when using step size of 1%. Therefore, tumor models that are more sophisticated with a gradient of HU can be designed by increasing the number of sub-volumes with different infill densities.

A very low error difference (in the range of µm) was observed when comparing the 3D/3D and 2D/3D registration performance between homogenous and heterogenous tumor models. We assume the reason for this is the effect of projection imaging; the heterogeneity induces only small changes in the radiological properties because of the small tumor volume in comparison to large volume of the remaining tissue compartments along the projection of the torso. The results showed that homogenous tumors with realistic HU are sufficient for validation of 2D/3D registration algorithms. However, the heterogenous phantom design, as proposed in this study, can be used by other research groups to investigate whether such complexity would be required for an exact verification and validation of their algorithms. A possible example could be the verification of advanced image segmentation algorithms. Depending on the algorithm, sharp edges of an object play a vital role in the performance of segmentation. Here, simulating realistic HU distributions in the tumor phantom can be crucial when validating segmentation methods [17, 18]. Our results showed a good agreement between different radiodensity values of the proposed phantom compared to the initial patient tumor CT scan. Using the Collision Detection approach, the resulting volume overlap was 96.72 % for the tumor model 1 and 78.15% for the tumor model 2. The observed error for the tumor model 2 can be attributed to the print resolution as the finer tumor geometry was not replicated in the smaller samples. Secondly, the threshold-based segmentation of the CT to generate digital models for comparison was pixelated, resulting in surface mismatch. As a future perspective of this study, we would like to examine some modern 3D printers which support the combination of different filament materials within one 3D print, this can help providing an additional option to modulate the radiological properties.

### 5) Conclusion

This research investigated the construction, radiological properties and geometrical accuracy of additive manufacturing-based tumor phantoms. This study lays the foundation of designing realistic and reliable tumor phantoms, which can be used for validation of different imaging algorithms in medical imaging area.

**Acknowledgment**

This work has been supported by ACMIT – Austrian Center for Medical Innovation and Technology, which is funded within the scope of the COMET program and funded by Austrian BMVIT and BMWFW and the governments of Lower Austria and Tyrol.

[1] N. Kiarashi, A.C. Nolte, G.M. Sturgeon, W.P. Segars, S.V. Ghate, L.W. Nolte et al. Development of realistic physical breast phantoms matched to virtual breast phantoms based on human subject data. Med. Phys 2015; 42: 4116–4126. doi: 10.1118/1.4919771.

[2] S Hatamikia et al., Additively manufactured patient-specific anthropomorphic thorax phantom with realistic radiation attenuation properties, Front. Bioeng. Biotechnol 2020; 8: 385. https://doi.org/10.3389/fbioe.2020.00385.

[3] R. Mayer, P. Liacouras, A. Thomas, M. Kang, L. Lin, C.B. Simone, 3D printer generated thorax phantom with mobile tumor for radiation dosimetry. Rev Sci Instrum 2015; 86: 074301. doi: 10.1063/1.4923294.

[4] Ehler ED, Barney BM, Higgins PD, Dusenbery E. Patient specific 3D printed phantom for IMRT quality assurance. Phys. Med. Biol. 2014; 59: 5763–73. doi: 10.1088/0031-9155/59/19/5763.

[5] Sepideh Hatamikia, Gernot Kronreif, Alexander Unger, Gunpreet Oberoi, Laszlo Jaksa, et al. 3D printed patient-specific thorax phantom with realistic heterogenous bone radiopacity using filament printer technology, Z Med Phys, 2022 (in press). https://doi.org/10.1016/j.zemedi.2022.02.001.

[6] V. Filippou, C. Tsoumpas, Recent advances on the development of phantoms using 3D printing for imaging with CT, MRI, PET, SPECT, and ultrasound. Med Phys 2018; 45: 740-760, 10.1002/mp.13058.

[7] P. Homolka, M. Figl, A. Wartak, M. Glanzer, M. Dünkelmeyer, A. Hojreh, et al. Design of a head phantom produced on a 3D rapid prototyping printer and comparison with a RANDO and 3 M lucite head phantom in eye dosimetry applications. Phys. Med. Biol 2017; 62: 3158-3174, 10.1088/1361-6560/aa602c

[8] N.I. Niebuhr, W. Johnen, T. Güldaglar, A. Runz, G. Echner, P. Mann, et al. Technical note: Radiological properties of tissue surrogates used in a multimodality deformable pelvic phantom for mr-guided radiotherapy. Med. Phy 43 2016; 43: 908-916, 10.1118/1.4939874.

[9] R. Tino, A. Yeo, M. Leary, M. Brandt, T. Kron. A systematic review on 3D-printed imaging and dosimetry phantoms in radiation therapy Technol. Cancer. Res. Treat. 2019; 18. 10.1177/1533033819870208.

[10] Hazelaar C, van Eijnatten M, Dahele M, Wolff J, Forouzanfar T, Slotman B, Verbakel WFAR. Using 3D printing techniques to create an anthropomorphic thorax phantom for medical imaging purposes. Med. Phys 2018; 45 :92-100. doi: 10.1002/mp.12644.

[11] Kostiukhina N, Georg D, Rollet S, Kuess P, Sipaj A, Andrzejewski P, Furtado H, Rausch I, Lechner W, Steiner E, Kertész H, Knäusl B. Advanced Radiation DOSimetry phantom (ARDOS): a versatile breathing phantom for 4D radiation therapy and medical imaging. Phys. Med. Biol 2017; 62: 8136–8153. doi: 10.1088/1361-6560/aa86ea.

[12] M. Sramek, Y, Shi, E. Quintanilla, X, Wu, A. Ponukumati et al. Tumor phantom for training and research in transoral surgery. Lary Inves Otolary 2020; 16: 677-682. doi: 10.1002/lio2.426.

[13] O.L. Dancewicz et al, Radiological properties of 3D printed materials in kilovoltage and megavoltage photon beams, Phys Medica 2017; 38: 111-118. doi: 10.1016/j.ejmp.2017.05.051.

[14] J. Solc et al., Tissue-equivalence of 3D-printed plastics for medical phantoms in radiology, J Instru 2018; 13. https://doi.org/10.1088/1748-0221/13/09/P09018.

[15] H. Furtado, E. Steiner, M. Stock, D. Georg, W. Birkfellner, Real-time 2D/3D registration using kV-MV image pairs for tumor motion tracking in image guided radiotherapy. Acta. Oncol 2013; 52: 1464–1471. doi: 10.3109/0284186x.2013.814152.

[16] Wolfgang Birkfellner1, Michael Figl, Hugo Furtado, Andreas Renner, Sepideh Hatamikia and Johann Hummel, Multi-Modality Imaging: A Software Fusion and Image-Guided Therapy Perspective. Front Phys 2018; 6: 66. https://doi.org/10.3389/fphy.2018.00066.

[17] C. Militello, L. Rundo, M. Dimaco, A. Orlando, R. Woitek, et al., 3D DCE-MRI Radiomic Analysis for Malignant Lesion Prediction in Breast Cancer Patients. Acad Radio 2021 (in press). https://doi.org/10.1016/j.acra.2021.08.024.

[18] N. Sharma, L.M. Aggarwal Automated medical image segmentation techniques. J Med Phys 2010; 35: 3-14. 10.4103/0971-6203.58777